\documentstyle[aps,pra,preprint,graphicx]{revtex}

\begin{document}

 \draft

\title{
Extended Gaussian wave packet dynamics }

\author{B.M.~Garraway}

\address{
      Sussex Centre for Optical and Atomic Physics,\\
      School of Chemistry, Physics, and Environmental Sciences,\\
      University of Sussex,
      Falmer, Brighton, BN1 9QJ,
      England
}
\date{\today.}
\maketitle
\begin{abstract}

  We examine an extension to the theory of Gaussian wave packet dynamics in
  a one-dimensional potential by means of a sequence of time dependent
  displacement and squeezing transformations. Exact expressions for the
  quantum dynamics are found, and relationships are explored between the
  squeezed system, Gaussian wave packet dynamics, the time dependent
  harmonic oscillator, and wave packet dynamics in a Gauss-Hermite
  basis. Expressions are given for the matrix elements of the potential in
  some simple cases. Several examples are given, including the propagation
  of a non-Gaussian initial state in a Morse potential.

\end{abstract}
\pacs{03.65.Ca, 31.70.Hq, 31.15.Qg, 82.40.Js}

\section{Introduction}
\label{sintro}

Wave packet dynamics has exposed interesting new phenomena in several
fields. In femto-chemistry \cite{femto,rpp} we are now able to time-resolve
chemical processes and also observe effects such as the breakup and revival
of wave packets \cite{averbukh}. In atom optics wave packets are used to
model matter waves \cite{atomoptics}, and electron wave packets are seen in
the dynamics of Rydberg atoms \cite{stroud}. The numerical modelling of wave
packet dynamics has been achieved by a number of methods
\cite{numerical,rpp}, but one of the earliest approaches was by Heller
\cite{heller} who simply used the Ansatz of a time dependent Gaussian wave
packet. This Gaussian approach is very useful, but it is usually an
approximation and can be quite wrong, for example at turning points. Several
improvements have been made: for example, the method of generalised Gaussian
wave packets \cite{hellercmplx} used complex classical trajectories for
Gaussian wave packets, and the hybrid method \cite{hellerhybrid} used an
expansion in terms of a grid of Gaussian wave packets.

The idea of using a time dependent harmonic (i.e.\ Gauss-Hermite) basis, in
the context of wave packet propagation, was put forward by Lee and Heller
\cite{leeheller} and Coalson and Karplus \cite{coalson82}. The basis was
chosen so that the lowest eigenstate matches the Heller Gaussian wave
packet, but with the inclusion of a complete set of basis states the
modelling can be performed accurately. This approach was generalised to
multi-dimensional systems by Lee \cite{lee86}. However, several
possibilities for using a Gauss-Hermite basis exist: parameters for the
dynamic basis were treated in a variational method by Kay \cite{kay89} and
by Kucar and Meyer \cite{kucar89}, more recently the phase space picture was
explored by M{\o}ller and Henriksen \cite{moller96a}, and the use of a
Gauss-Hermite basis with a variational treatment has been expanded by
Billing \cite{billing97,billing99a,adhikari99} to examine non-adiabatic
transitions and corrections to classical path equations.

The approach that is used in this paper is similar, in principle, to the
time dependent Gauss-Hermite basis of Refs.\ \cite{leeheller} and
\cite{coalson82}; in each case the basis follows the Heller 
Gaussian wave packet.  However, the focus here is on the evolution operator
and transformations associated with the time-dependent basis which becomes
both ``displaced'' and ``squeezed''.  That is, unlike previous extensions to
Gaussian wave packet dynamics, the system Hamiltonian is transformed by
displacement and squeezing in a way that removes {\em all} operator
dependence which is quadratic or less.  The result is an evolution equation
which depends on a time dependent `residual potential' which is based on the
original (one-dimensional) 
potential with harmonic terms removed. This means that if the
higher order derivatives of the potential are small, the system evolution
will change relatively slowly, which allows a rapid numerical integration
(of a set of ordinary differential equations).  The method, which is in
principle exact, is then similar to a time-dependent perturbation theory in
a time dependent basis or interaction picture.  Indeed, it can be developed
as a perturbation theory in the higher order derivatives of the system
potential about the classical motion.

Because the evolution operator is
found, the possibility exists for applying this extended Gaussian wave
packet method to initial states that are not Gaussian.
The evolution operator also allows us to use
the displacement and squeezing
transformations to find explicit matrix elements of the residual
potential using standard operator algebra.

In section~\ref{sdisp} of this paper we set up the problem and perform a
basis shift according to the classical dynamics.
Section~\ref{sgauss} examines Gaussian wave packet dynamics in this
displaced basis, and in the original basis by two different approaches.
The relationship between the two approaches is established.
In section~\ref{ssqu} we establish the squeezing transformation necessary to 
map the time evolution of a Gaussian wave packet from its initial state.
By using the same transformation for another change of basis we can find the 
equations for corrections to Gaussian wave packet dynamics.
These equations are expressed in a Fock basis in section~\ref{sfock}, where
we also compare our results to the Gauss-Hermite basis.  Some examples of
useful matrix elements for potentials are given in section~\ref{sresidual},
and in section~\ref{sapplied} the results are applied to several different
problems.

\section{Displaced basis}
\label{sdisp}

\subsection{Scaling of the problem}
\label{ssscale}

The problem we wish to describe is the one dimensional problem of a wave
packet in a potential described by the Hamiltonian
\begin{equation}
   {\widetilde H} = \frac{ {\widetilde p}^2}{2m} 
 + {\widetilde U}( {\widetilde x} )  
 \;.
\end{equation}
The position and momentum co-ordinates have been denoted by ${\widetilde x}$ 
and ${\widetilde p}$ to distinguish them from scaled quantities that will 
shortly be introduced.
The initial wave packet will be taken to be a Gaussian one. This is
not essential, but it will simplify the treatment that will
follow. 
The key feature is that a length scale, characteristic of the initial wave
packet defines the width of the harmonic oscillator basis that we will use.
For a Gaussian initial wave packet 
\begin{equation}
  \Psi_0({\widetilde x}) = \frac{1}{ (2\pi\sigma_0^2)^{1/4} } \exp \left[ 
              -\frac{ ({\widetilde x}-{\widetilde x}_0)^2 }{ 4\sigma_0^2 } 
              + i \frac{ {\widetilde p}_0 {\widetilde x}}{\hbar} \right]
 \;,
 \label{wpinit}
\end{equation}
which has a width $\sigma_0$ and is located at ${\widetilde x}_0$ with
momentum $ {\widetilde p}_0$.

We will now adopt a scaling of the problem such that we use the operators
\begin{eqnarray}
  \hat x = \frac{{\widetilde x}}{  \sqrt{2} \sigma_0 } \nonumber\\
  \hat p = \frac{\sqrt{2} \sigma_0 {\widetilde p} }{\hbar}
 \label{scale}
\end{eqnarray}
which have the commutator $[\hat x,\hat p]=i$. The Schr\"odinger equation 
then reduces to
\begin{equation}
   i\frac{\partial \Psi}{\partial t} = 
  \left[ \frac{ \hat p^2}{ 2 } + U(\hat x) \right] \Psi
 \;,
 \label{schr}
\end{equation}
where we use scaled time and energy,
\begin{equation}
   t =    {\widetilde \omega}_0   {\widetilde t}
\end{equation}
\begin{equation}
   U =   {\widetilde U }  /(
           \hbar  {\widetilde \omega}_0 )
 \;.
\end{equation}
 The frequency ${\widetilde \omega}_0$ is determined by the width of the
initial wave packet. It is the frequency  of the harmonic oscillator for
which the wave function (\ref{wpinit}) is a ground state wave function,
\begin{equation}
 {\widetilde \omega}_0 = \frac{ \hbar }{ 2 m \sigma_0^2 }
 \;.
\end{equation}
In terms of the scaled quantities the initial wave function is now
\begin{equation}
  \Psi_0(x) = \frac{1}{ \pi^{1/4} } \exp \left[ 
              -\frac{ (x-x_0)^2 }{ 2 } 
              + i p_0 x \right]
 \;,
 \label{wpsinit}
\end{equation}
when we use the appropriately scaled $x_0$ and $p_0$.

\subsection{Local expansion of the potential}
\label{ssexpan}

The motion of a Gaussian wave packet in a harmonic potential is exactly
solvable, even when the potential is time dependent, and we will use this to 
define the local basis for the wave packet. That is, the potential function
$U(x)$ will be expanded to second order about the the position of the wave
packet. The dynamics of a Gaussian wave packet in this harmonic potential
will be determined, and these dynamics will be used to define the basis for
the full (non-Gaussian) wave packet dynamics.

The wave packet (\ref{wpsinit}) is located at the position $x_0$, which will 
in general be time dependent; we then take its location to be given by
$x_0(t)$. If we expand the potential about this point we
obtain, 
\begin{equation}
   U(x) =  U(x_0) + U'(x_0) ( x - x_0 ) +  \frac{ U''(x_0) }{2} 
  ( x - x_0)^2   + U_R(x, x_0)
 \;,
 \label{uexpand}
\end{equation}
where the spatial derivatives are indicated with the primes. The terms
in the expansion
are not explicitly time dependent; they vary only with time through the
position $x_0(t)$. 
The potential function $U_R(x, x_0)$ is the residual potential found after
making the harmonic expansion. That is, it contains the higher order, cubic
and above, terms in the expansion. The residual potential will play a central 
role in the non-Gaussian dynamics of the wave packet, and  Eq.\ 
(\ref{uexpand}) serves as its definition.

\subsection{Displacement of the basis}
\label{ssdisp}

In the following, we will make two basis changes in order to match the
Gaussian part of the wave packet dynamics.  The first basis transformation
will be a displacement to remove the linear term in $x$ from the potential in
Eq.\ (\ref{uexpand}). The necessary displacement is simply $x_0(t)$ in space
and a momentum $p_0(t)$, such that the wave function is shifted to the
origin. The new wave function will be
\begin{equation}
    \psi_d(x,t) = \hat D(-\beta(t)) \Psi(x,t) \;,
 \label{wpsid}
\end{equation}
i.e.\ $\Psi(x,t)  = \hat D(\beta(t))  \psi_d(x,t)$,
where $\hat D(\beta(t))$ is the time dependent displacement operator
\begin{equation}
  \hat D(\beta(t)) = 
 \hat D^{-1}(-\beta(t)) = 
    \exp\left(   i p_0(t) \hat x    - i x_0(t) \hat p 
    \right) \;,
 \label{dbeta}
\end{equation}
with
\begin{equation}
 \beta = \frac{ x_0 + i p_0 }{\sqrt{2}} \;.
 \label{beta}
\end{equation}
The potential $U(x)$ in the  Schr\"odinger equation (\ref{schr}) will become
transformed as $  D^{-1}(\beta(t))  U(\hat x)  \hat D(\beta(t)) = U(x_0 +\hat
x)$ and we will then use the expansion in Eq.\ (\ref{uexpand}). 

The requirement to remove the linear $\hat x$ term (and linear $\hat p$
term) from the potential means that after inserting Eq.\ (\ref{wpsid}) in
the Schr\"odinger equation (\ref{schr}) we obtain the conditions:
\begin{eqnarray}
    x_0' &=& p_0(t) \label{classx}\\
    p_0' &=& -U'(x_0(t)) \label{classp}
\end{eqnarray}
which are, of course, the classical equations of motion.
With these conditions the linear term in $\hat x$ is lost and the
Schr\"odinger equation now reads
\begin{equation}
   i\frac{\partial \psi_d(x,t)}{\partial t} = 
  \left[ \frac{ \hat p^2}{ 2 } + U(x_0(t)) 
   -\frac{1}{2} U'(x_0(t))  x_0(t) 
   +\frac{ U''(x_0(t)) }{2} \hat x^2  + U_R(x_0 + \hat x, x_0)
  \right] \psi_d(x,t)
 \;.
 \label{ud}
\end{equation}
The non-operator parts of Eq.\ (\ref{ud}) are easily removed with a time
dependent phase factor 
\begin{equation}
  \phi_U(t) = \int^t \left[ 
                U(x_0(t'))   -  \frac{1}{2} x_0(t') U'(x_0(t'))
         \right] dt'
           =  \int^t \left[ \frac{1}{2} x_0(t') p'_0(t') +  U(x_0(t'))
         \right] dt'
 \;,
 \label{phase}
\end{equation}
where for the second form we have used Eq.\ (\ref{classp}). Then if we define
the displaced wave function with a phase shift as
\begin{equation}
  \psi_{dp}(x,t) = e^{i\phi_U(t)}\psi_d(x,t) =e^{i\phi_U(t)} \hat D(-\beta(t)) 
                   \Psi(x,t)
 \;,
 \label{wpsidp}
\end{equation}
we obtain  the Schr\"odinger equation
\begin{equation}
   i\frac{\partial \psi_{dp}(x,t)}{\partial t} = 
  \left[ \frac{ \hat p^2}{ 2 }
   +\frac{ U''(x_0(t)) }{2} \hat x^2  + U_R(x_0 + \hat x, x_0)
  \right] \psi_{dp}(x,t)
 \;.
 \label{schrdp}
\end{equation}
Basis displacement has been of interest in the study of quantum state
diffusion (QSD) \cite{ianp} where the non-linear dynamics create wave
packet localisation. By using a displaced basis a reduction in computational
effort is gained. However, in quantum state diffusion there is no strong
motive for going to the next step of squeezing the basis because QSD
localised wave packets all have the same size. In ordinary Schr\"odinger
wave packet dynamics wave packets can change their widths enormously making
basis squeezing desirable.

\section{Gaussian wave packet dynamics}
\label{sgauss}

\subsection{Heller's approach}

For completeness we include here an outline of standard Gaussian wave 
packet dynamics. 
Heller  started with the
Ansatz \cite{heller} 
\begin{equation}
    \Psi_{GWP} = \exp\left[ i \alpha(t) (x-x_0(t))^2 + i p_0(t)(x-x_0(t)) +
               i\gamma(t)\right] 
 \label{gwp}
\end{equation}
in the original basis [here we use the scaled basis of Eq.\ (\ref{scale})].
The normalisation is included in the time dependent complex parameter
$\gamma(t)$,
and
Heller introduced the parameter $\alpha(t)$ which characterises (the
reciprocal of) the width of the Gaussian wave packet. The position $x_0$ and 
momentum $p_0$ of the wave packet obey the classical equations of motion,
exactly as in Eqs.\ (\ref{classx}) and (\ref{classp}). By substituting the
Gaussian wave packet into the Schr\"odinger equation (\ref{schr}) with the
truncated potential
\begin{equation}
   U(x) \sim  U(x_0) + U'(x_0) (x - x_0 ) +  \frac{ U''(x_0) }{2} 
  (x - x_0)^2  \;,
  \label{utrunc}
\end{equation}
we can show that
\begin{eqnarray}
\alpha'  &=& -2\alpha^2 - U''(x_0)/2  \nonumber\\
 \gamma' &=& i\alpha + p_0 x_0' - E
 \label{gwpeq}
\end{eqnarray}
where $E$ is the classical energy $p_0^2/2 + U(x_0)$.

Thus the dynamics of an approximate Gaussian wave packet are completely
defined
by solving the differential equations (\ref{classx}), (\ref{classp}), and
(\ref{gwpeq}). The result is approximate because Eq.\ (\ref{utrunc}) is an
approximation to Eq.\ (\ref{uexpand}).

\subsection{Gaussian wave packets in the displaced basis}

In order to establish some notation, and motivate the squeezing
transformation in section~\ref{ssqu}, this section gives an overview
of Gaussian wave
packet dynamics as found in the displaced basis of Eq.\ (\ref{wpsid}).
Thus starting with Eq.\ (\ref{schrdp}), 
we again neglect the residual potential $U_R$ to obtain
\begin{equation}
   i\frac{\partial \psi_{s}(x,t)}{\partial t} = 
  \left[ \frac{ \hat p^2}{ 2 }
   +\frac{ k(t) }{2} \hat x^2 
  \right] \psi_{s}(x,t) \;,
 \label{schrs}
\end{equation}
where $k(t)=U''(x_0(t))$ is a time dependent spring constant.
This Schr\"odinger problem  does have a known time
dependent `ground' state solution (see, for example, Ref.\ \cite{glauber}),
which is not a stationary state, because of the time dependence
in $k(t)$. The `ground' state solution can be
formulated in terms of local classical trajectories.
Using some of the notation of Ref.\ \cite{glauber} we define a quantity
$\epsilon(t)$, through the equation
\begin{equation}
  \epsilon''(t) = - k(t) \epsilon(t)  \;,
 \label{epsdef}
\end{equation}
which would make $\epsilon$ follow the classical trajectory of a point close
to the centre of the wave packet. For $\epsilon$ we have the following,
complex, initial conditions
\begin{eqnarray}
 \epsilon(0) &=& 1 \nonumber\\
 \epsilon'(0) &=& i 
 \label{epsinit}
\end{eqnarray}
so that we have a time independent Wronskian with the value
\begin{equation}
  {\cal W} = \epsilon' \epsilon^\ast - \epsilon \epsilon'\!\,^\ast
             = 2i
 \;.
 \label{w}
\end{equation}
Then the ground state wave function takes the 
form \cite{glauber}
\begin{equation}
 \psi_{s}(x,t) = \frac{1}{ \pi^{1/4}\sqrt{\epsilon(t)}}   \exp
        \left[ \frac{ i\epsilon'(t) }{ 2\epsilon(t) } x^2  \right]
 \;,
 \label{psis}
\end{equation}
as may be verified by substitution into Eq.\ (\ref{schrs}).  We note that
the time dependent width of this wave packet is
$1/\surd[2$Im$(\epsilon'/\epsilon)]$ which is found to be
$|\epsilon|/\sqrt{2}$ on using the Wronskian (\ref{w}).  The ground state
wave function (\ref{psis}) is identical to the Heller Gaussian
wave packet if we transform it back to the original basis using the inverse
of Eq.\ (\ref{wpsidp}). That is,
\begin{equation}
 \Psi_{GWP} = e^{-i\phi_U(t)} \hat D(\beta(t))  \psi_{s}(x,t) \;.
\end{equation}
If we perform the displacement of the wave packet we obtain Eq.\ (\ref{gwp})
with the identifications:
\begin{eqnarray}
  \alpha(t) &=&  \epsilon'(t) / [ 2 \epsilon(t) ]  \nonumber\\
  \gamma(t) &=& x_0p_0/2 - \phi_U(t) + g_n
\end{eqnarray}
where $g_n$ is a complex term arising from the normalisation of
Eq.\ (\ref{psis}), i.e.
\begin{equation}
   \exp( - i g_n ) = \pi^{1/4}\sqrt{\epsilon(t)}
 \;.
\end{equation}
Thus we see that the Heller Gaussian wave packet corresponds to the
evolution of the `ground' state of the time dependent harmonic oscillator
(\ref{schrs}).

\section{Time dependent squeezed basis}
\label{ssqu}

In order to go beyond the Gaussian wave packet approximation we need to take
into account the non-Gaussian behaviour introduced by the residual potential
$U_R$. This could, of course, be achieved in any reasonable basis, but in
order to take advantage of the power of Gaussian wave packet dynamics, which
is often such a good approximation to the time evolution, it makes sense to
use a time dependent basis which matches the Gaussian wave packet. To find
this basis it is not enough to use the displaced basis of
section~\ref{ssdisp}; we must also squeeze the basis to match the (time
dependent) width of the Gaussian wave packet as well as its location. Thus,
in the same way that we use a displacement operator to remove the linear
dependence of the Hamiltonian on $\hat x$ in section~\ref{ssdisp}, we will
here use a squeezing transformation to remove {\em all} quadratic ($\hat x$
and $\hat p$) terms from the Hamiltonian, thereby leaving the naked
dependence on the residual potential $U_R$.  
This will ensure that in the special case where the residual potential is
zero, $U_R=0$, the transformed wave function is stationary. In this case the 
displacement and squeezing transformations will map an initial Gaussian wave 
packet onto Heller's moving Gaussian wave packet (\ref{gwp}).
To remove the quadratic operator dependence we denote the
squeezing transformation as $\hat U_s$ and define a wave function $ \psi_{sdp}$
in the squeezed and displaced basis as
\begin{equation}
  \psi_{sdp}(x,t) = \hat U_s^{-1}\psi_{dp}(x,t) =
\hat U_s^{-1} e^{i\phi_U(t)}\psi_d(x,t) =\hat U_s^{-1} e^{i\phi_U(t)} \hat
D^{-1}(\beta(t)) \Psi(x,t)
 \;.
 \label{wpsisdp}
\end{equation}
Writing Eq.\ (\ref{schrdp}) as
\begin{equation}
   i\frac{\partial \psi_{dp}(x,t)}{\partial t} = 
  \left[ H_s(t)   +   U_R(x_0+\hat x, x_0)
  \right] \psi_{dp}(x,t)
 \;,
 \label{schrdp2}
\end{equation}
we may substitute for $\psi_{dp}$ from Eq.\ (\ref{wpsisdp}) to obtain
\begin{equation}
     i  \hat U_s^{-1} \frac{ \partial  \hat U_s }{\partial t}  \psi_{sdp}
       +   i\frac{\partial \psi_{sdp}}{\partial t} 
            =
          \hat U_s^{-1} \left[  H_s +   U_R(x_0+\hat x, x_0) \right] \hat
U_s \psi_{sdp}
 \;.
 \label{uspsieqn}
\end{equation}
The term $H_s$ contains all the quadratic operator dependence and can be
removed if the operator $ \hat U_s $ obeys
 \begin{equation}
     i  \hat U_s^{-1} \frac{ \partial  \hat U_s }{\partial t} =
          \hat U_s^{-1}  H_s  \hat U_s
 \;.
 \label{useqn} 
\end{equation}

To determine $\hat U_s$
we will start in the basis of the displaced harmonic oscillator, using
the annihilation and creation operators
\begin{eqnarray}
 \hat a         &=& \frac{ \hat x + i \hat p}{\sqrt{2}} \nonumber\\
 \hat a^\dagger &=& \frac{ \hat x - i \hat p}{\sqrt{2}}  
 \label{aadagdef}
\end{eqnarray}
so that the Hamiltonian of Eq.\ (\ref{schrs}) 
[$H_s$ in Eq.\ (\ref{schrdp2})] becomes
\begin{equation}
    H_s = \frac{1}{4\epsilon}\left[ 2(\epsilon-\epsilon'')\hat N
         - (\epsilon+\epsilon'')( \hat a^2 + \hat a^{\dagger 2})  \right]
 \;,
 \label{hs}
\end{equation}
 where
\begin{equation}
\hat N = \frac{ \hat a^\dagger \hat a + \hat a \hat  a^\dagger }{2}
       = \hat a^\dagger \hat a  + 1/2
 \;.
 \label{Ndeff}
\end{equation}

After some consideration we express the unitary operator $\hat U_s$ in the form
\begin{equation}
   \hat U_s = \hat S(\xi) e^{-i\hat N\theta}  \;,
 \label{USdeff}
\end{equation}
where $\hat S(\xi)$ is the usual squeezing operator \cite{squop}
\begin{equation}
    \hat S(\xi)  = \exp\left( -\frac{\xi}{2} \hat a^{\dagger 2}
                             +\frac{\xi^\ast}{2} \hat a^2 \right)
 \;.
 \label{sqdeff}
\end{equation}
Then using the standard expressions for the action of the squeeze operator 
\cite{squop}, and the
phase shifting properties of $e^{-i\hat N\theta}  $, the operator $ 
\hat U_s $ will transform the annihilation and creation operators as
\begin{eqnarray}
\hat U_s^{-1}\hat a   \hat U_s        &=&  \hat a e^{-i\theta} \cosh r 
                      - \hat a^\dagger e^{i(\phi+\theta)} \sinh r 
 \nonumber \\
\hat U_s^{-1} \hat a^\dagger \hat U_s  &=&  \hat a^\dagger e^{i\theta} 
            \cosh r    - \hat a e^{-i(\phi+\theta)} \sinh r
  \label{sqtdef}
\end{eqnarray}
where 
\begin{equation}
      \xi  =  re^{i\phi} \;.
  \label{xidef}
\end{equation}
Explicit expressions for $r$, $\phi$, and $\theta$, will be found by 
substitution in Eq.\ (\ref{useqn}). However, 
to determine $ \hat U_s^{-1} \frac{ \partial  \hat U_s }{\partial t}$ in
Eq.\ (\ref{useqn}) from the Ansatz (\ref{USdeff}) we need to differentiate
the exponential operator $\hat S$ with respect to the time dependence of its 
parameters. This is accomplished by first disentangling the operator,
i.e.\ by using the relation \cite{squent}
\begin{equation}
   \hat S(\xi)  =  
       \exp\left( -\frac{1}{2} e^{ i\phi}\tanh r \hat a^{\dagger 2}\right)
       \exp\left[ - \ln(\cosh r) \hat N \right]
       \exp\left(  \frac{1}{2} e^{-i\phi}\tanh r \hat a^2 \right)
  \label{disent}
\end{equation}
and then differentiating. It is then necessary to pull the 
non-exponential factors containing
$ \hat a $ and $\hat a^\dagger$ to one side, and re-entangle the squeeze
operator before a comparison can be made between both sides of Eq.\
(\ref{useqn}). 
Some details of this calculation are presented in
Appendix \ref{snasty}. 
The final results for the squeezing, and phase,
parameters are
\begin{equation}
       \tanh r = 
        \frac{\left| \epsilon + i \epsilon' \right|}{
              \left| \epsilon - i \epsilon' \right|}
 \label{tanhr}
\end{equation}
\begin{equation}
	e^{i\phi} = - 
       \frac{ (\epsilon + i \epsilon') }{\left| \epsilon + i \epsilon' \right|}
       \frac{\left| \epsilon - i \epsilon' \right|}{ (\epsilon - i \epsilon') }
 \label{phi}
\end{equation}
\begin{equation}
	e^{i\theta} = 
       \frac{ (\epsilon - i \epsilon') }{\left| \epsilon - i \epsilon'
\right|}
  \;.
 \label{theta}
\end{equation}
A few additional relations between $r, \phi$, and $\theta$ can be found in
Appendix \ref{snasty}.

In this way Eq.\ (\ref{useqn}) is solved and we are left with
the
residual potential $ U_R(x_0+\hat x, x_0)$ in Eq.\ (\ref{uspsieqn}).
Because $ U_R(x_0+\hat x, x_0)$ depends 
on the operator $\hat x$ it will be transformed under the squeezing
transformation (\ref{USdeff}). Simply by using Eq.\ (\ref{aadagdef})
with Eq.\ (\ref{sqtdef}) we find that
\begin{equation}
\hat U_s^{-1} \hat x \hat U_s  = \mbox{Re}[\epsilon(t)] \hat x 
                                +\mbox{Im}[\epsilon(t)] \hat p 
 \;.
\label{xs}
\end{equation}
Then we obtain from Eq.\ (\ref{uspsieqn}) the ``Schr\"odinger equation'' 
\begin{equation}
   i\frac{\partial \psi_{sdp}}{\partial t}(x,t) = 
       U_R\left( x_0+  \mbox{Re}[\epsilon(t)] \hat x 
             +\mbox{Im}[\epsilon(t)] \hat p \;  ,\; x_0\right)
   \psi_{sdp}(x,t)
 \;,
 \label{schrsdp}
\end{equation}
which is one of our key results. It describes the evolution of a wave
function entirely in terms of a ``potential'' with  higher than quadratic
behaviour, i.e.\ in terms of the residual potential $U_R$.

We note that neglect of $U_R$ in Eq.\ (\ref{schrsdp}) returns us to 
Gaussian wave packet dynamics. In this case $\psi_{sdp}$ is stationary
and we thus find from Eq.\ (\ref{wpsisdp}) that
\begin{equation}
 \Psi_{GWP}(x,t) =   e^{-i\phi_U(t)}  \hat D(\beta(t))   \hat U_s(t)
     \hat U_s^{-1}(0)  \hat D^{-1}(\beta(0))
  \Psi(x,t=0)
 \;.
\end{equation}
This expresses the Heller Gaussian wave packet in terms of a sequence of
displacement and squeezing transformations, and would allow us, for example, 
to propagate a non-Gaussian wave packet in the same way as the Heller
Gaussian wave packet is propagated in time.

\section{Fock state implementation}
\label{sfock}

Our result so far, Eq.\ (\ref{schrsdp}), describes corrections to Gaussian
wave packet dynamics, but is hard to implement because of the appearance of
the operator $\hat p$ throughout the transformed $U_R$.  However, it is
amenable to treatment in a Fock basis. If we return to the operators $\hat
a$ and $\hat a^\dagger$ in Eq.\ (\ref{aadagdef}) we can write
\begin{equation}
\hat U_s^{-1} \hat x \hat U_s  = \frac{ \epsilon^\ast(t)\hat a + \epsilon(t) 
\hat a^\dagger}{\sqrt{2}} 
 \;,                      
 \label{xsfock}
\end{equation}
so that the  Eq.\ (\ref{schrsdp}) becomes
\begin{equation}
   i\frac{\partial \psi_{sdp}}{\partial t}(x,t) = 
       U_R\left(  x_0(t) + \frac{ \epsilon^\ast(t)\hat a + \epsilon(t) 
\hat a^\dagger}{\sqrt{2}}   
      \;  ,\; x_0(t) \right)
   \psi_{sdp}(x,t)
  \;.
 \label{schrsdpfock}
\end{equation}
If we now solve Eq.\ (\ref{schrsdpfock}) in a Fock basis we obtain the
non-Gaussian wave packet dynamics, i.e.\ the extended Gaussian wave packet,
in the basis defined by the motion of the Gaussian part of the wave packet.
To do this we expand the wave function in the Fock basis (of states labelled 
$|n\rangle$), defined by the Gaussian
wave packet ground state 
\begin{equation}
   \psi_{sdp} (x,t) = \sum_{n} a_n(t)   |n\rangle
 \;.
 \label{psifock}
\end{equation}
 Then the equation of
motion becomes
\begin{equation}
i \frac{\partial a_n(t)}{\partial t} = \sum_m
      \left\langle n \left|
      U_R\left(  x_0+ \frac{ \epsilon^\ast \hat a + \epsilon
\hat a^\dagger}{\sqrt{2}}   
      \;  ,\; x_0\right)     \right|m\right\rangle
a_m(t)
 \;.
 \label{fock}
\end{equation}
The equations which have now to be solved depend on the form of $U_R$ and
its matrix elements. In section~\ref{sresidual} we will look at some
specific functional forms for the residual potential in order to determine
explicit expressions for Eq.\ (\ref{fock}) when matrix elements are taken.

To determine the spatial wave function in the original (scaled) basis and in
terms of the coefficients $a_n$ we need to express Eq.\ (\ref{psifock}) in
the original spatial basis by using the transformation (\ref{wpsisdp}). That
is, using also Eq.\ (\ref{USdeff}),
\begin{equation}
   \Psi(x,t) = e^{  -i\phi_U(t) }   \sum_{n=0}^\infty a_n
    e^{ -i(n+1/2) \theta }
    \left\langle x \left|
      \hat D(\beta(t))  \hat S(\xi)
     \right|n\right\rangle
   \;.
 \label{spatial}
\end{equation}
We then utilize the spatial
distribution of squeezed displaced Fock states \cite{moller96b} and after some
calculation obtain
\begin{eqnarray}
  \Psi(x,t) = && 
   \frac{1}{ \pi^{1/4} \sqrt{\epsilon}}
   \exp\left\{i\left[ \frac{\epsilon'}{2\epsilon} (x-x_0)^2 +xp_0-x_0p_0/2
     -\phi_U(t)  \right]\right\}
   \nonumber\\ && \times
   \sum_{n=0}^\infty a_n
  \left(\frac{|\epsilon|}{\sqrt{2}\epsilon}\right)^n
   \frac{1}{\sqrt{n!}}
    H_n\left(\frac{x-x_0}{|\epsilon|}\right)
  \;.
 \label{hermite}
\end{eqnarray}
The values of $x_0(t), p_0(t), \epsilon(t), \epsilon'(t)$ and $a_n(t)$
can be used to determine the spatial wave function.
This result for $\Psi$ can be compared to the Ansatz employed as the
starting point of the analysis used in Refs.\ 
\cite{leeheller,coalson82,kay89,kucar89,billing99a}, which each use a
Gauss-Hermite basis. Refs.\ \cite{kay89,kucar89,billing99a} all use a
variational method where the parameters of the basis depend on the wave
function. Refs.\ \cite{leeheller,coalson82} use a basis similar to 
Eq.\ (\ref{hermite}), but since they chose
the simplest kind of basis related to the 
Heller Gaussian wave packet (\ref{gwp}) for $n=0$, the expansion 
used differs from Eq.\ (\ref{hermite}) by phase factors. The
$n$th term in  Eq.\ (\ref{hermite}) has an additional phase of
  $(|\epsilon|/\epsilon)^n$.
Similar phase factors, are absent in variational treatments, for example from
Billing's Ansatz \cite{billing99a} (see also
Appendix~\ref{appbilling}). The $n$ dependence of this phase factor means
that some quadratic operator dependence is still present in the 
equations for the
amplitudes of the $n$th Fock state. However, the variational methods try to
optimise the wave packet trajectory---a process
we do not consider here which may compensate. 
Also note that while Eq.\ (\ref{fock}) requires the
evaluation of matrix elements of the residual potential, similar matrix
elements for the full potential are required in Ref.\
\cite{billing99a}. Again, some analytical approaches to these matrix
elements are given in section~\ref{sresidual}.

Finally, we expect to perform a numerical integration of the various
equations to determine the wave packet dynamics of our particular system.
The equations which have to be numerically integrated are: (i) the classical
equations of motion Eqs.\ (\ref{classx}), (\ref{classp}), (ii) the
$\epsilon$ equation (\ref{epsdef}), which may be split into two complex
first order differential equations, or four real first order equations, and
(iii) the amplitudes of the corrections in Eqs.\ (\ref{fock}). For a basis
size of $N$ states, including the lowest energy state, this amounts to
$6+2N$ real, first order, linear differential equations.  
The initial conditions are specified by the initial position and momentum of
the wave packet, the initial conditions for $\epsilon$ [Eq.\
(\ref{epsinit})], and, in the case of a Gaussian initial wave packet, $a_0=1,
a_{n \ne 0} =0$.
Then once the various matrix elements in Eq.\ (\ref{fock}) have been set up,
typically involving some finite sums (see
the next section), the numerical integration can be done in a
straightforward way. Note that, unlike Ref.\ \cite{leeheller}, it is not
necessary to use finite difference methods on nearby trajectories, which can
result in a reduction of numerical effort.

\section{Matrix elements of the residual potential}
\label{sresidual}

\subsection{Exponential terms in $U_R$}

We consider a term in $U_R$ of the form
\begin{equation}
  U_\beta( x ) =  \exp(-\beta_U x )  
 \;,
\end{equation} 
where $\beta_U$ is a constant characterising the potential.
This term might arise from consideration of a Morse potential
and in that case there would be two
exponential terms like this one (see Section \ref{ssmorse1}).
Then, following Eq.\ (\ref{schrsdpfock}), we will need to evaluate
\begin{equation}
U_{\beta n m} =  \left\langle n \left|   U_\beta \left(
\frac{ \epsilon^\ast(t)\hat a + \epsilon(t)  \hat a^\dagger}{\sqrt{2}}
  \right)       \right|m\right\rangle
  =
 \left\langle n \left| 
   \exp\left[-\beta_U  \frac{ \epsilon^\ast(t)\hat a + \epsilon(t) 
\hat a^\dagger}{\sqrt{2}}  \right]
  \right|m\right\rangle
 \;.
 \label{ur.exp}
\end{equation}
 We
 start by disentangling the operators in the
exponential,
\begin{equation}
 \exp\left[-\beta_U  \frac{ \epsilon^\ast(t)\hat a + \epsilon(t) 
\hat a^\dagger}{\sqrt{2}}  \right]
  = 
 \exp( -\beta_U\epsilon\hat a^\dagger /\sqrt{2} )
\exp( -\beta_U\epsilon^\ast\hat a /\sqrt{2} )
\exp( \beta_U^2 |\epsilon|^2 / 4 )
 \;.
\end{equation}
We can then proceed several ways. For example, writing the exponentials as a 
power series and using 
\begin{equation}
 \hat a^k |m\rangle   = \sqrt\frac{m!}{(m-k)!} |m-k\rangle
 \;,
\end{equation}  
we obtain 
\begin{equation}
\exp( -\beta_U\epsilon^\ast\hat a /\sqrt{2} )   |m\rangle   =
  \sum_{k=0}^m\frac{  (-\beta_U\epsilon^\ast/\sqrt{2} )^k
            }{k!}\sqrt{\frac{ m!}{(m-k)!} } |m-k\rangle
 \;,
\end{equation}  
with a similar expression for $ \langle n |\exp(
-\beta_U\epsilon\hat a^\dagger /\sqrt{2} )$. Putting both of these expressions
together we obtain the finite sums
\begin{equation}
U_{\beta n m}
 = \left\{ \begin{array}{ll}  {\displaystyle
 \sqrt{n!m!}\;e^{|\eta|^2/2} \;(-\eta)^\Delta  \;
  \sum_{k=0}^m\frac{   |\eta|^{2k} }{ k! (k+\Delta )! ( m-k)! }  }
    & \;\;\;\;\;\;\;   n\ge m   \\
  {\displaystyle
 \sqrt{n!m!}\;e^{|\eta|^2/2} \;(-\eta^\ast)^{-\Delta}  \;
  \sum_{k=0}^n\frac{   |\eta|^{2k} }{ k! (k-\Delta )! ( n-k)! }  }
    & \;\;\;\;\;\;\;   n\le m   
  \end{array}\right.
 \;,
 \label{ubeta}
\end{equation}
where
\begin{equation}
  \Delta = n-m
\end{equation} 
and
\begin{equation}
  \eta = \beta_U \epsilon /\sqrt{2}
 \;.
\end{equation}

\subsection{Power terms in $U_R$}
\label{sspower}

We consider a term in $U_R$ of the form
\begin{equation}
  U_q( x ) =  x^q  
 \;,
\end{equation} 
where $q$ is an integer. Terms like this could arise in any Taylor series
expansion of a potential.
Again, following Eq.\ (\ref{schrsdpfock}), we will need to evaluate
\begin{equation}
U_{q n m} = 
\left\langle n \left|   U_q \left(
\frac{ \epsilon^\ast(t)\hat a + \epsilon(t)  \hat a^\dagger}{\sqrt{2}}
  \right)       \right|m\right\rangle
  =
\left\langle n \left|
   \left[  \frac{ \epsilon^\ast(t)\hat a + \epsilon(t) 
\hat a^\dagger}{\sqrt{2}}  \right]^q
  \right|m\right\rangle
 \;.
 \label{ur.pow}
\end{equation}
One way to proceed is to recognise that the exponential operator in Eq.\
(\ref{ur.exp}) can be written as
\begin{equation}
\exp\left[-\beta_U  \frac{ \epsilon^\ast(t)\hat a + \epsilon(t) 
\hat a^\dagger}{\sqrt{2}}  \right]
 =
\sum_{q=0}^\infty \frac{(-\beta_U)^q}{q!}
   \left[  \frac{ \epsilon^\ast(t)\hat a + \epsilon(t) 
  \hat a^\dagger}{\sqrt{2}}  \right]^q
 \;,
\end{equation}
so that the coefficient of $(-\beta_U)^q/q!$ in Eq.\ (\ref{ubeta})
will lead to the matrix element (\ref{ur.pow}).
Thus, after expanding the exponential in Eq.\ (\ref{ubeta}), and writing 
$\epsilon$ in terms of its modulus and phase as $\epsilon=|\epsilon|
 e^{i\theta_\epsilon}$ we find
\begin{equation}
U_{q n m}
 = \left\{ \begin{array}{l}  {\displaystyle
 q!
\sqrt{n!m!}\;  \frac{|\epsilon|^q
                 e^{i\Delta\theta_\epsilon}}{2^{q-\Delta/2}}
   \;  
  \sum_{k=0}^%
{{\scriptsize \mbox{min}}(m,(q-\Delta)/2)}
  \frac{  2^k}{  k!
   (k+\Delta)! (m-k)! [  (q-\Delta)/2 -k ]!  } }
     \\  \hspace{9cm}
 n\ge m ,\;q -\Delta \ge 0  \\
  {\displaystyle
 q!
\sqrt{n!m!}\;  \frac{|\epsilon|^q
                 e^{i\Delta\theta_\epsilon}}{2^{q+\Delta/2}}
   \;  
  \sum_{k=0}^%
{{\scriptsize \mbox{min}}(n,(q+\Delta)/2)}
  \frac{  2^k}{  k!
   (k-\Delta)! (n-k)! [  (q+\Delta)/2 -k ]!  } }
     \\  \hspace{9cm}    n\le m   ,\;q +\Delta \ge 0
  \end{array}\right.
 \;,
 \label{uq}
\end{equation}
where for non-zero matrix elements we must have $q$ even if $\Delta = n-m$
is even, or $q$ odd if $\Delta = n-m$ is odd. 
We must also have $q\ge 0$.
Note that these coefficients
have a very simple dependence on $|\epsilon|$ and $\theta_\epsilon$. This
means that if the matrix elements are calculated on a computer, the (finite) 
sums do not need to be completely re-evaluated as $\epsilon$ changes in time.

We can write down a few examples for small $q$, which can also be determined 
by explicitly expanding Eq.\ (\ref{ur.pow}). 
For reference, some results are shown in Table~\ref{tuqlow}
for the non-zero matrix elements when ($n$=0, 1, 2, 3...)

\subsection{Using a Taylor series for the potential}

We can now use the results of section~\ref{sspower} to determine the
transformed residual potential when we expand the potential as a Taylor
series about the classical position of the wave packet,
\begin{equation}
   U(x) = \sum_{k=0}^\infty  \frac{ U^{(k)}(x_0) }{ k!}  (x-x_0)^k
 \;,
 \label{ljexpand}
\end{equation}
with $U^{(k)}(x)$ as the $k$th derivative of the potential.
Then the residual potential is simply given by
\begin{equation}
    U_R(x, x_0) = \sum_{k=3}^\infty  \frac{ U^{(k)}(x_0) }{ k!}  (x-x_0)^k
 \;,
 \label{ljexpandr}
\end{equation}
and the transformed residual potential, as in Eq.\ (\ref{fock}), is
\begin{equation}
     U_R\left(  x_0+ \frac{ \epsilon^\ast \hat a + \epsilon
     \hat a^\dagger}{\sqrt{2}}    \;  ,\; x_0\right) 
     =
    \sum_{k=3}^\infty  \frac{ U^{(k)}(x_0)  }{ k!}  
  \left[
 \frac{
\epsilon^\ast(t)\hat a + \epsilon(t)  \hat a^\dagger}{\sqrt{2}} 
 \right]^k
 \;.
 \label{ljur}
\end{equation}
The matrix elements of the last part of Eq.\ (\ref{ljur}), needed in
Eq.\ (\ref{fock}), can be determined by using the result found in Eq.\
(\ref{uq}),
\begin{equation}
      \left\langle n \left|
      U_R\left(  x_0+ \frac{ \epsilon^\ast \hat a + \epsilon
      \hat a^\dagger}{\sqrt{2}}   
      \;  ,\; x_0\right)     \right|m\right\rangle
  =
    \sum_{k=3}^\infty  \frac{ U^{(k)}(x_0)  }{ k!}  
    U_{k n m}
  \;.
 \label{ljur2}
\end{equation}

For example, the Lennard-Jones potential
\begin{equation}
   U_{LJ}(x) = \frac{ C_1}{ x^{12} } - \frac{C_2}{  x^{6}  }
 \;,
 \label{lj.def}
\end{equation}
may be expanded as
\begin{equation}
    U_{LJ}^{(k)}(x) = \frac{ (-1)^k }{x^k} 
     \left(  C_1 \frac{ (k+11)! }{ 11!}\frac{1}{ x^{12}}  
        -  C_2 \frac{ (k+5)! }{ 5! } \frac{1}{ x^{6}}  
     \right)
 \;,
 \label{ljdiff}
\end{equation}
leading to
\begin{equation}
      \left\langle n \left|
      U_R\left(  x_0+ \frac{ \epsilon^\ast \hat a + \epsilon
      \hat a^\dagger}{\sqrt{2}}   
      \;  ,\; x_0\right)     \right|m\right\rangle
  =
    \sum_{k=3}^\infty  \frac{ (-1)^k  }{x_0^k k!}
  \left(  C_1 \frac{ (k+11)! }{ 11!}\frac{1}{ x_0^{12}}  
        -  C_2 \frac{ (k+5)! }{ 5! } \frac{1}{ x_0^{6}}  
     \right)
    U_{k n m}
  \;.
 \label{ljur3}
\end{equation}

\section{Examples}
\label{sapplied}

\subsection{Application to an exponential potential: N\lowercase{a}I}
\label{ssnai}

In this section we will apply the techniques developed in the previous
sections to an example with a potential energy that varies exponentially
with distance. Such a problem could arise in atom optics where mirrors made
from evanescent waves \cite{evmirror}, or a periodic magnetisation
\cite{magmirror}, have an exponential dependence on distance from the
mirror. However, the example we will consider here arises in a diatomic
molecule where the energy of the molecule depends on the inter-nuclear
separation (denoted by $R$ here).

\begin{figure}[h]
 \begin{center}
  \includegraphics[scale=1.0]{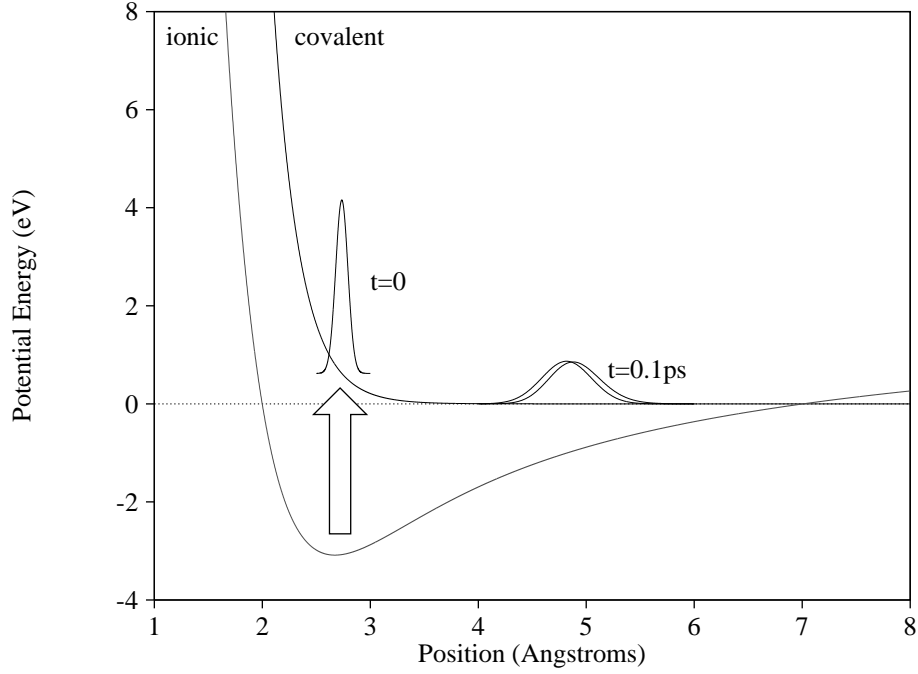}
 \end{center}
\caption[x]{\label{fig1}
 Relevant potentials for the NaI molecule with superimposed wave
packets. Expressions for the ionic and covalent potentials are given in
Eqs.\ (\ref{ionic}) and (\ref{covalent}). The wave packets shown for $t=0$
and $t=0.1$ ps have been scaled. 
The wave packet at $t=0$ has been promoted from the ionic potential to the
covalent potential by an ultra-short pulse.
The two wave packets shown at  $t=0.1$ ps
arise from the Gaussian wave packet approximation (left) and the extended
Gaussian wave packet approach (right). These two wave packets are shown more 
clearly in Fig.\ \ref{fig2}.
}
\end{figure}

Wave packet dynamics in the NaI molecule have been much described 
\cite{nairefs,choi,engel}, both theoretically and experimentally. 
Figure~\ref{fig1} shows the essential elements of the system. We consider
two potential energy curves: one ionic and one covalent. The unexcited
system is ionic, but at large separations of the two atoms there is an
attractive Coulomb force which results in the level crossing near to 7 \AA\
separation.  For this system we may use the potentials of 
Refs.\ \cite{faist} and
\cite{vanveen} as, for example, used in Refs.\ \cite{choi,engel} with
\begin{equation}
V_1(R) = A_1 \exp( - \beta_1 (R-R_0))
\label{covalent}
\end{equation}
for the covalent surface and
\begin{eqnarray}
V_2(R) =&& (A_2+(B_2/R)^8) \exp(-R/\rho)-e^2/R-e^2(\lambda_+
+\lambda_-)/2R^4 
\nonumber\\ &&
-C_2/R^6 - 2 e^2 \lambda_+ \lambda_-/R^7 + \Delta E
\label{ionic}
\end{eqnarray}
for the ionic surface.
The centrifugal term in the dynamics is neglected.
The values for the constants in Eqs.\ (\ref{covalent}) and (\ref{ionic})
are taken from \cite{engel} (see Table \ref{tab1}).

In experiments on NaI (see Refs.\ \cite{nairefs}) the ground state wave
packet on the ionic surface is subjected to an ultra short laser pulse which
places a wave packet on the covalent surface as indicated for $t=0$ on Fig.\
\ref{fig1}. On the covalent surface the wave packet is not in an equilibrium
position, and so it it starts to move towards the crossing point. At the
crossing the wave packet divides into two pieces, and the subsequent
oscillations of the wave packet in the upper adiabatic surface have been
much discussed \cite{nairefs,choi,engel}.

Here we will suppose the laser pulse is sufficiently short that its
amplitude can be considered to be a delta function. In that case the wave
packet on the covalent surface at $t=0$ is simply proportional to the ground
state wave function \cite{choi,rpp}. As a result the ionic potential
(\ref{ionic}) will only serve to define the initial wave packet appearing on
the covalent potential: 
we take it here to be a Gaussian with a
width of
0.056 \AA.  This narrow packet spreads quite rapidly, making this system of
interest to an extended Gaussian wave packet method. We concern ourselves
initially with the wave packet motion, prior to the crossing point, on the
exponential potential (\ref{covalent}).

With the potential given by Eq.\ (\ref{covalent}) the residual potential
[Eq.\ (\ref{uexpand})] is (we will now use $x$, rather than $R$, for the
co-ordinate to maintain the notation of sections \ref{sintro}-\ref{sfock})
\begin{equation}
 U_R(x, x_0)  = U(x_0 + x ) - U(x_0) - U'(x_0)x -U''(x_0)x^2/2 
 \;.
\end{equation}
Then when we move to the squeezed basis, $x$ becomes transformed as
$x \rightarrow x_0 + {\tilde{\tilde x}}$, with ${\tilde{\tilde x}}= 
\hat U_s^{-1} \hat x \hat U_s $ as in Eq.\ (\ref{xsfock}). When we
form the matrix elements, as in Eq.\ (\ref{fock}), we obtain
\begin{eqnarray}
      \left\langle n \left|
      U_R\left(  x_0+ \frac{ \epsilon^\ast \hat a + \epsilon
      \hat a^\dagger}{\sqrt{2}}   
      \;  ,\; x_0\right)     \right|m\right\rangle
    &=& 
      \left\langle n \left|
      U_R\left(   {\tilde{\tilde x}}    \;  ,\; x_0\right)   
       \right|m\right\rangle
 \nonumber\\ &=&
  A_1 \exp( - \beta_1 (x_0(t)-R_0)) \Bigl[
 <n|\exp(-\beta_1 {\tilde{\tilde x}})|m> - \delta_{n,m} 
  \nonumber\\ && 
  + \beta_1 <n|{\tilde{\tilde x}}|m> 
  - \beta_1^2<n|{\tilde{\tilde x}}^2|m> /2 \Bigr]
 \;.
  \label{naiur}
\end{eqnarray}
Then the exponential term in Eq.\ (\ref{naiur}) may be evaluated by using
Eq.\ (\ref{ubeta}), and the two power terms may be evaluated with the $q=1,2$
results from Table~\ref{tuqlow}, i.e.
\begin{eqnarray}
   \left\langle n \left|
      U_R\left(   {\tilde{\tilde x}}    \;  ,\; x_0\right)   
       \right|m\right\rangle
    &=& 
  A_1 \exp( - \beta_1 (x_0(t)-R_0)) 
  \nonumber\\ && \times
    \Bigl[
    U_{\beta_1 n m} - \delta_{n,m} 
  + \beta_1  U_{1 n m}
  - \beta_1^2 U_{2 nm} /2 \Bigr]
 \;.
  \label{naiur2}
\end{eqnarray}
With these matrix elements for the residual
potential, the equations to be numerically integrated are: Eqs.\
(\ref{classx}), (\ref{classp}), and (\ref{epsdef}).

The results for the NaI molecule may be seen in Fig.\ \ref{fig1} as the two
wave packets on the right of the figure. The full, extended wave packet is
furthest to the right, and the simple Gaussian wave packet approximation,
Eq.\ (\ref{gwp}), is seen just slightly to the left.  The wave packets have
been derived from the amplitudes $a_n$ by simply combining the $a_n$ with
the spatial representation of the harmonic oscillator eigenfunctions, as in
Eq.\ (\ref{hermite}), and then squaring the result.

\begin{figure}[h]
 \begin{center}
  \includegraphics[scale=1.0]{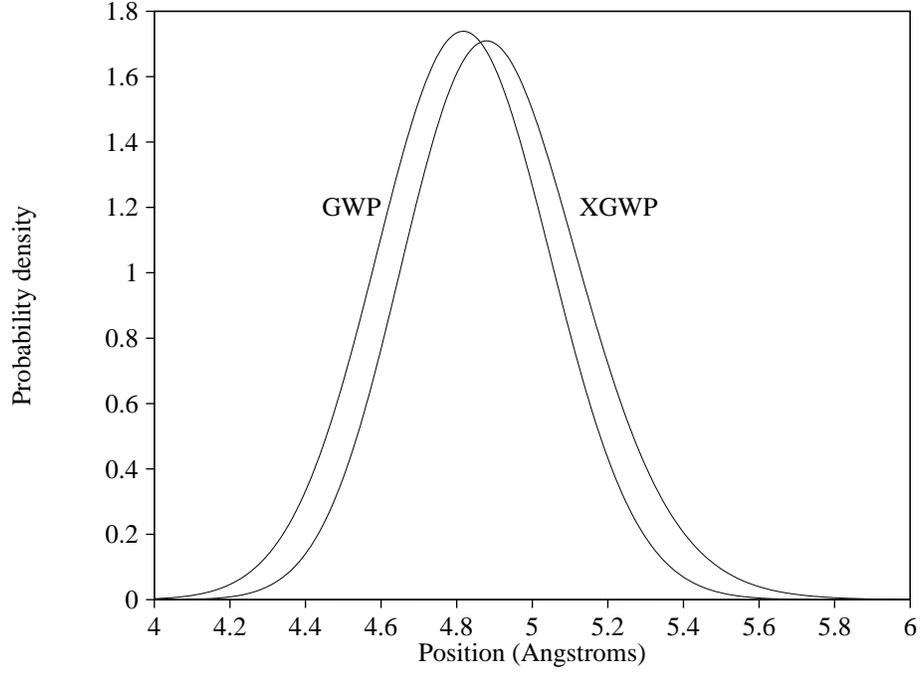}
 \end{center}
\caption[x]{\label{fig2}
 Wave packet evolution in NaI after a time of 0.1 ps. The wave packet on the 
left (GWP) is from the Gaussian wave packet approximation, Eq.\
(\ref{gwp}). The wave packet on the right (XGWP) is the
extended Gaussian wave packet from Eqs.\ (\ref{classx}), (\ref{classp}),
(\ref{epsdef}), and Eqs.\ (\ref{fock}) with six basis states.
}
\end{figure}

The same two wave packets are shown on a larger scale in Fig.\ \ref{fig2}. 
The extended Gaussian wave packet (XGWP) is on the right, and as well as
being displaced from the Gaussian wave packet it does not have 
a Gaussian shape because its base is skewed slightly to right.
These results have been tested against a standard numerical integration of
the Schr\"odinger equation using a split step fast Fourier transform method
(see, for example, Ref.\ \cite{rpp} for a description of the method).
The number of basis states used for the extended Gaussian wave packet in
Fig.\ \ref{fig2} was just six. However, for this example it is found that
just four basis states are sufficient for a reasonable approximation to the
wave packet.

\subsection{Morse potential with Gaussian and non-Gaussian initial state}
\label{ssmorse1}

We consider next wave packet dynamics in a Morse potential model with
a potential
\begin{equation}
  U_M(x) = D\left[ 1 - \exp(-\beta_M(x-x_M)) \right]^2  \;.
\label{umorse}
\end{equation}
Because this potential can be expressed as the sum of two exponentials and a
constant, we can use the results of Eqs.\ (\ref{naiur},\ref{naiur2}) (for the
exponential potential) to find the matrix elements of the residual
potential in the squeezed basis:
\begin{eqnarray}
   \left\langle n \left|
      U_R\left(   {\tilde{\tilde x}}    \;  ,\; x_0\right)   
       \right|m\right\rangle
    &=& 
  D \Bigl\{  
     \exp\left[ -2  \beta_M (x_0(t)-x_M) \right] 
     \left[  U_{2\beta~n m} - \delta_{n,m} \right]
  \nonumber\\ && %
       -2  \exp\left[ - \beta_M (x_0(t)-x_M)\right]
     \left[  U_{\beta~n m} - \delta_{n,m} \right]
     \Bigr\}
  \nonumber\\ && %
     -U_M'( x_0(t) ) U_{1 n m}   -U_M''( x_0(t) ) U_{2 n m}
%  \nonumber\\ && \times %
%
%           utmp = d*(  -2.0d0*exp(a*(xc-x))*(unm1-idelta) 
%     +                + exp(2.0d0*a*(xc-x))*(unm2-idelta)
%     +    )
%     +           -umorse(1, var, x)*uqnm1 
%     +      - 0.5d0*umorse(2, var, x)*uqnm2 
%
%
 \;,
  \label{morseur2}
\end{eqnarray}
where $ U_{\beta~n m}$ and $ U_{2\beta~n m}$ are given by Eq.\
(\ref{ubeta}), $ U_{1 n m}$ and  $  U_{2 n m}$ are given in
Table~\ref{tuqlow},
and $U_M'$ and $U_M''$ are the first and second derivatives of the Morse
potential (\ref{umorse}) which will be evaluated at the classical position
$ x_0(t)$.

\begin{figure}[h]
 \begin{center}
  \includegraphics[scale=1.0]{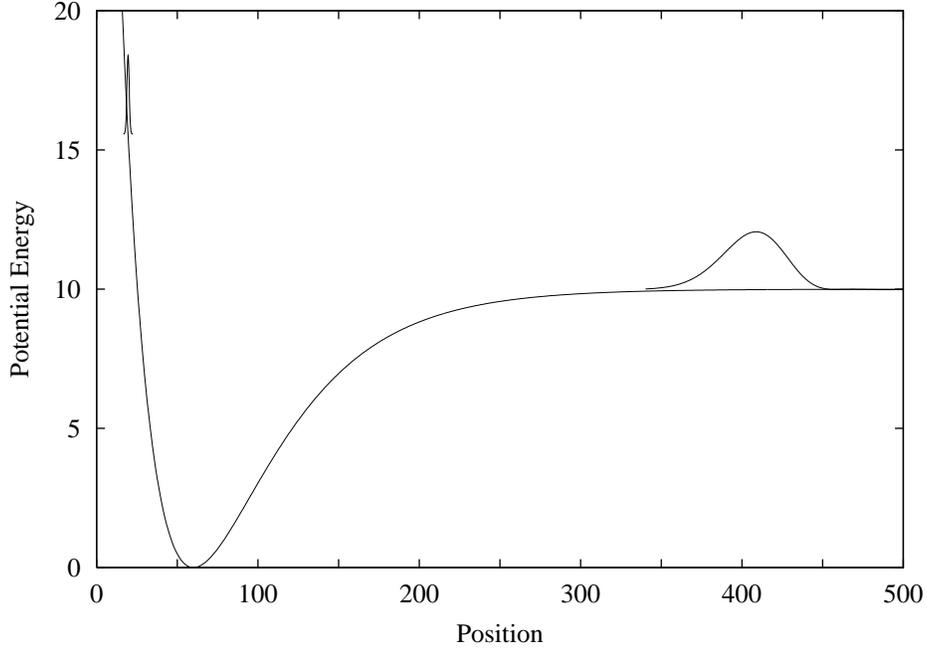}
 \end{center}
\caption[x]{\label{fig3} Morse potential showing the initial Gaussian
wave packet (left) and the final time evolved wave packet (right)
which is shown in more detail in Fig.\ \ref{fig4}. The initial and
final wave packets are shown together here to show the dramatic change
of width; a situation where the extended Gaussian wave packet approach
may be appropriate. The initial wave packet is located at $x=19.5$
(with a width of $1/\sqrt{2}$ in scaled units) and the final wave
packet is shown for $t=70.0$ in scaled units (i.e.\ $m=1, \hbar=1$).
The Morse potential, Eq.\ (\ref{umorse}), has parameters $D=10,
\beta_M=0.02$, and $x_M=60$.  }
\end{figure}

\begin{figure}[h]
 \begin{center}
  \includegraphics[scale=1.0]{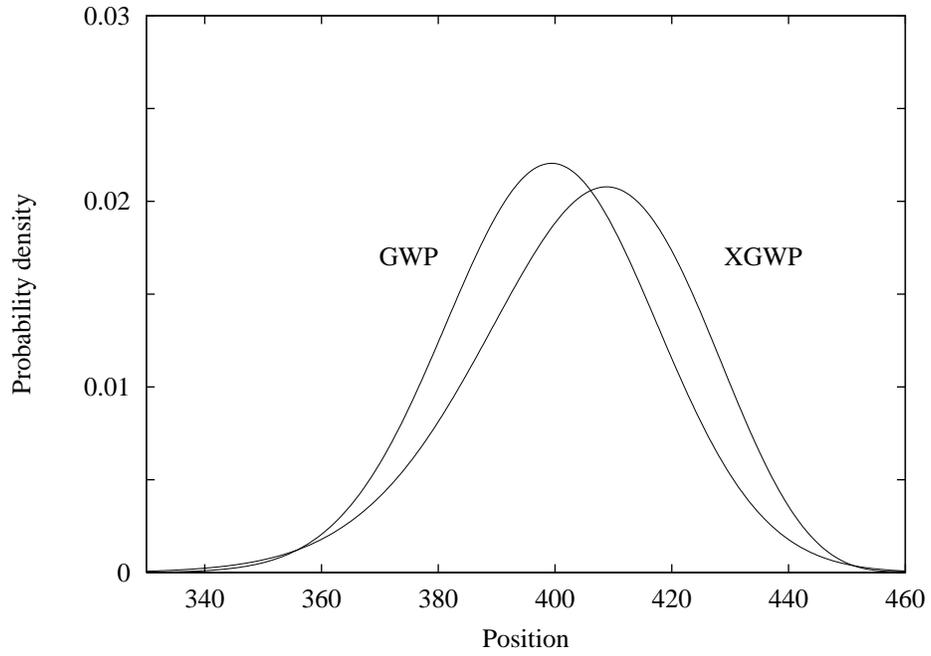}
 \end{center}
\caption[x]{\label{fig4}
Final wave packet, determined with 20 basis states (XGWP), for the evolution
on the Morse potential in Fig.\ \ref{fig3}. The Gaussian wave packet (GWP)
appears skewed to the left. The time elapsed since the initial state is 70.0 
in scaled units. }
\end{figure}

The parameters for the potential (see Fig.\ \ref{fig3}) have been
chosen so that the initial wave packet broadens considerably during
the time evolution. The scaled units of Section \ref{ssscale} are used
(equivalent to $m=1,\hbar=1$) with an initial wave packet width of
$1/\sqrt{2}$.  The results shown in Figures \ref{fig3} and \ref{fig4}
show how the wave packet develops an asymmetry. In Fig.\ \ref{fig4}
the final wave packet (computed with 20 basis states) is compared to
the Gaussian wave packet (\ref{gwp}). We see that the top of the
extended Gaussian wave packet is shifted to the right in the Figure,
and appears distinctly non-Gaussian when compared to the Gaussian wave
packet. The 20 basis states used are sufficient to obtain convergence.

\begin{figure}[h]
 \begin{center}
  \includegraphics[scale=1.0]{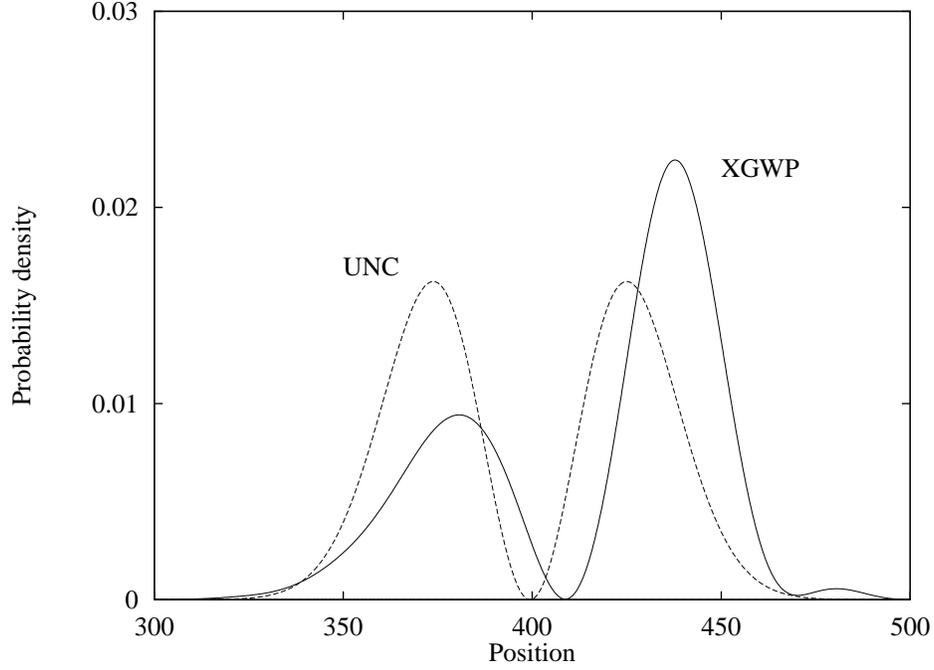}
 \end{center}
\caption[x]{\label{fig5}
Final wave packets, determined at $t=70$, for an initial $n=1$ wave packet
located at $x=19.5$ in the potential shown in Fig.\ \ref{fig3}. The solid
curve (XGWP) shows the extended Gaussian wave packet calculated in a basis
of 20 states. The dashed curve (UNC) shows the final wave packet when there
is no coupling between the basis states, and the initial wave packet only
changes its location, width, and momentum.  }
\end{figure}

It was briefly mentioned at the end of Section \ref{ssqu} that the squeezing 
and displacement transformations could be used to propagate a non-Gaussian
wave packet. This is straightforwardly done in the Fock basis of Section
\ref{sfock} where it is simply a question of assigning the initial
amplitudes $a_n$ in Eq.\ (\ref{psifock}). Figure \ref{fig5} shows the
results of such a case where the initial state was chosen such that
$a_1=1$ (with the remaining amplitudes set to zero) corresponding to the
first excited state of a harmonic oscillator. In this case the spatial wave
packet has the form:
\begin{equation}
   \Psi(x) = \frac{ 2(x-x_0) }{\pi^{1/4} } \exp[-(x-x_0)^2 /2 ]
    \;.
\end{equation}
This initial state is
propagated in the same potential shown in Fig.\ \ref{fig3} and for the same
time as the Gaussian initial wave packet was propagated in Fig.\
\ref{fig4}. The curve marked XGWP in Fig.\ \ref{fig5} shows the result of
the extended Gaussian wave packet propagation with 20 basis states. The
dashed curve (marked UNC) shows the wave packet that results when there is
no coupling from the $n=1$ Fock state in the dynamic basis. In this case the
final wave packet has the same form as the initial wave packet (i.e.\ it is
still characterised by $n=1$) but the width, position and momentum have all
changed. This means that the curve marked UNC amounts to the same kind of
approximation to the actual final wave packet (XGWP) in Fig.\ \ref{fig5} as
the Gaussian wave packet in Fig.\ \ref{fig4} is to the actual (XGWP) wave
packet there. For  Fig.\ \ref{fig5}, we see that, as for the Gaussian
initial wave packet in Fig.\ \ref{fig4}, it is important to have the
coupling of the residual potential.

The ability of the extended Gaussian wave packet method to be used for such
non-Gaussian initial states can clearly increase the applicability of this
type of method. Not only can the ground states of anharmonic potentials be
propagated, but we could also propagate a thermal wave packet. In this case
we would separately propagate the thermally populated vibrational states and 
then add (with thermal weightings) the final probability distributions.

\section{conclusion}
\label{sconc}

In this paper we have seen a description of wave packet dynamics in
terms of a time dependent Gaussian basis. Explicit expressions have been
found for the displacement and squeezing parameters that describe the basis, 
and the displacement and squeezing transformations have been used to
determine analytic expressions for matrix elements of simple forms of
potentials. By expanding a potential as a Taylor series about the classical
trajectory it is possible to use the analytic expressions for the matrix
elements (in a truncated expansion) for almost any reasonable potential
[as in Eq.\ (\ref{ljur2})]. As an example, the extended Gaussian
wave packet method was applied to the dissociation of NaI.

The extended Gaussian wave packet (XGWP) method is good for wave
packet evolution where the packet remains close to a Gaussian one, and
the method is especially appropriate if there are large changes in
scale during the motion (as found in the examples treated in
section~\ref{sapplied}). In these cases we can expect the XGWP method
to be faster than a numerical grid propagation method, and more
accurate than a plain Gaussian wave packet method. Whether it is
faster, or how much faster the XGWP method is, will depend on a
particular situation. In the case of the example treated in
section~\ref{ssnai}, a numerical split operator FFT method was
found to be roughly a thousand times slower than the XGWP method.

Finally, we should note that while the idea of the Gauss-Hermite basis has
been exploited by a number of authors, in different ways, the emphasis in
this paper has been on the transformations involved. 
Only 1D results have been presented, and it is not clear if the method
extends easily to more degrees of freedom.
The method may not be so good
for collision processes where the development of large asymmetries in
the wave packet can result in large excitation of the squeezed basis.
However, the method does seem appropriate for dissociative processes where
there are large changes in scale, and the potential does not change on a
length scale much smaller than the wave packet. The extended Gaussian wave
packet method presented here can also be used to propagate non-Gaussian wave 
packets.
Finally, although other Gauss-Hermite methods may have a similar numerical
performance, it is hoped that the analytic results given here may
give useful insights in the future.

\acknowledgments

This work was supported by the United Kingdom Engineering and Physical
Sciences Research Council.

\appendix

\section{Determination of \lowercase{$r$}, $\phi$, and $\theta$}
\label{snasty}

If we differentiate Eq.\ (\ref{disent}) we obtain
\begin{equation}
\frac{\partial}{\partial t} \hat S(\xi) =
  -\frac{A'}{2} \hat a^{\dagger 2} \hat S(\xi)  
  +\frac{A'\!\,^\ast}{2} \hat S(\xi)
\hat a^2 
     - B' e^{-A  \hat a^{\dagger 2} /2 }  
          e^{-B \hat N} \hat N e^{A^\ast \hat a^2 /2 }
\label{a1}
\end{equation}
where
\begin{eqnarray}
   A &=&  e^{ i\phi}\tanh r \nonumber\\
   B &=& \ln(\cosh r) 
   \;.
\label{a2}
\end{eqnarray}
Then on shifting the non-exponential $\hat N$ term to the right we can form 
\begin{eqnarray}
i e^{i\hat N\theta}
\hat S(-\xi) \frac{\partial}{\partial t} \hat S(\xi) e^{-i\hat N\theta} 
 &=&
 -i  \frac{A'\cosh^2r}{2} \hat a^{\dagger 2} e^{2i\theta}
  \nonumber \\&& %
 +i\frac{1}{2}\left[ 2B'A^\ast + A'\!\,^\ast - A'\sinh^2r e^{-2i\phi}
\right] \hat a^2  e^{-2i\theta}
  \nonumber \\&& %
 + i \left[ A'e^{-i\phi} \sinh r\cosh r - B' \right] \hat N
\label{a3}
\end{eqnarray}
as will be required for the LHS of Eq.\ (\ref{useqn}). We will also need
\begin{equation}
 i e^{i\hat N\theta} \frac{\partial}{\partial t}  e^{-i\hat N\theta}
  = \hat N \theta'
\;.
\label{a4}
\end{equation}
Now if we let
\begin{eqnarray}
   y &=& \cosh r \nonumber\\
   z &=& e^{ i\phi}\sinh r
   \;,
\label{a5}
\end{eqnarray}
so that
\begin{eqnarray}
   A' &=& (z'y-zy')/y^2 \nonumber\\
   B' &=& y'/y
   \;,
\label{a6}
\end{eqnarray}
the LHS of Eq.\ (\ref{useqn}) can be written as 
 \begin{eqnarray}
     i  \hat U_s^{-1} \frac{ \partial  \hat U_s }{\partial t} 
 &=&
\frac{ -i \hat a^{\dagger 2} e^{2i\theta} }{2} (z'y - zy')
+ \frac{ i \hat a^{2} e^{-2i\theta} }{2} ( z'\!\,^\ast y - z^\ast y')
+ i\hat N (z'z^\ast - y' y -\theta' )
\;.
\label{a7}
\end{eqnarray}

According to Eq.\ (\ref{useqn}) $H_s$ will become squeezed, and on using
Eqs.\ (\ref{sqtdef}) we will find 
 \begin{eqnarray}
 \hat U_s^{-1}  H_s  \hat U_s
 &=&
  \frac{1}{4\epsilon}\left\{
-a^{\dagger 2} e^{2i\theta}\left[ 
   2(\epsilon-\epsilon'')yz + (\epsilon+\epsilon'')(y^2+ z^2)
   \right]
   \right.
 \nonumber \\&& %
-a^{2} e^{-2i\theta}\left[ 
   2(\epsilon-\epsilon'')yz^\ast + (\epsilon+\epsilon'')(y^2+ z^{\ast 2} )
   \right]
 \nonumber \\&& %
  \left.
 + 2 \hat N \left[
     (\epsilon-\epsilon'')( y^2 +|z|^2) + (\epsilon+\epsilon'')(yz+yz^\ast)
    \right]
    \right\}
\;.
\label{a8}
\end{eqnarray}
Then comparing Eq.\ (\ref{a8}) with Eq.\ (\ref{a7}), and inspecting the
coefficient of $\hat a^{\dagger 2}$ we find that
 \begin{eqnarray}
  \frac{\partial}{\partial t} (z/y) = \frac{\partial}{\partial t} A
  = -\frac{i}{2\epsilon}\left[
   \epsilon(A+1)^2 + \epsilon''(A-1)^2 \right]
\label{a9}
\end{eqnarray}
with the solution
 \begin{eqnarray}
A =	e^{i\phi} \tanh r  = -\frac{ \epsilon + i \epsilon' }{ 
                                   \epsilon - i \epsilon' }
\;.
\label{a10}
\end{eqnarray}
Then comparing Eq.\ (\ref{a8}) with Eq.\ (\ref{a7}), and inspecting the
coefficient of $\hat N$ we find, after some algebra, that
\begin{eqnarray}
 \theta' = -i \frac{\epsilon' -i\epsilon''}{ \epsilon -i\epsilon'}
\label{a11}
\end{eqnarray}
and
\begin{eqnarray}
 e^{i\theta} =  \frac{\epsilon -i\epsilon'}{| \epsilon -i\epsilon'|}
\;.
\label{a12}
\end{eqnarray}
It then follows that
\begin{equation}
       \cosh r = \frac{1}{2}( \epsilon - i \epsilon' )e^{-i\theta} 
               = \frac{1}{2}\left| \epsilon -i \epsilon' \right| 
 \;,
\label{a13}
\end{equation}
\begin{equation}
       \sinh r =-\frac{1}{2}( \epsilon + i \epsilon' )e^{-i(\theta+\phi)} 
               = \frac{1}{2}\left| \epsilon +i \epsilon' \right| 
 \;.
\label{a14}
\end{equation}

\section{}
\label{appbilling}

To connect Eqs. (\ref{fock}) and (\ref{hermite}) more closely to 
the work of Billing \cite{billing99a}
we would
define some coefficients
\begin{equation}
   c_n = (\epsilon^\ast/\epsilon)^{n/2 + 1/4}  a_n
   \label{bcoeff}
\end{equation}
and then insert Eq.\ (\ref{bcoeff}) into Eq.\ (\ref{fock}), using
\begin{equation}
   \frac{\partial}{\partial t} (\epsilon^\ast/\epsilon)^{n/2+ 1/4}
    = -i\frac{n}{|\epsilon|^2}  (\epsilon^\ast/\epsilon)^{n/2+ 1/4}
    \equiv  -2i\mbox{Im}(\alpha) n (\epsilon^\ast/\epsilon)^{n/2+ 1/4}
\end{equation}
to obtain,
\begin{eqnarray}
i \frac{\partial c_n(t)}{\partial t} = 
  &&
  \mbox{Im}(\alpha) (2n+1)  c_n(t)
  \nonumber \\ &&
   +
     \sum_m
      \left\langle n \left|
      U_R\left(  x_0+ \frac{ \epsilon^\ast(t)\hat a + \epsilon(t) 
\hat a^\dagger}{\sqrt{2}}   
      \;  ,\; x_0\right)     \right|m\right\rangle
 (\epsilon^\ast/\epsilon)^{(n-m)/2}
c_m(t)
 \label{bfock}
\end{eqnarray}
as the equation of motion for the coefficients $c_n$.

\begin{table}
\begin{tabular}{llll}
  &$n $&$m $&$ U_{qnm} $\\
\tableline
$q=1 $&$ n$&$ n+1 $&$ \epsilon^\ast \sqrt{\frac{n+1}{2}}   $\\
$    $&$ n+1$&$ n $&$ \epsilon  \sqrt{\frac{n+1}{2}}      $\\
\tableline
$q=2    $&$ n$&$ n $&$ |\epsilon|^2 \frac{n+1}{2}   $\\
$  $&$ n$&$ n+2 $&$ \epsilon^{\ast 2} \frac{\sqrt{(n+1)(n+2)}}{2}$\\
$  $&$ n+2$&$ n $&$ \epsilon^{     2} \frac{\sqrt{(n+1)(n+2)}}{2}$\\
\tableline
$q=3  $&$ n$&$ n+3 $&$ \frac{1}{2\sqrt{2}}  \epsilon^{\ast 3}
          \sqrt{(n+1)(n+2)(n+3)} $\\
$  $&$ n+3$&$ n $&$ \frac{1}{2\sqrt{2}}  \epsilon^{3}
          \sqrt{(n+1)(n+2)(n+3)} $\\
$  $&$ n$&$ n+1 $&$ \frac{3}{2\sqrt{2}}  \epsilon\epsilon^{\ast 2}
           (n+1)^{3/2}  $\\
$  $&$ n+1$&$ n $&$ \frac{3}{2\sqrt{2}}  \epsilon^{2}\epsilon^\ast
           (n+1)^{3/2} $
\end{tabular}

\caption[x]{\label{tuqlow}
 For the potential $x^q$, values of  $U_{qnm}$ are shown as determined
 from Eq.\ (\ref{uq}).
}
\end{table}

\begin{table}
\begin{tabular}{rdrd}
Ionic & & Covalent &\\
\tableline
$A_2$[eV] &2760                  &$A_1$ [eV] & 0.813    \\
$B_2$ [eV$^{1/8}$\AA]&2.398      &$\beta_1$ [\AA$^{-1}$]& 4.08   \\
$C_2$ [eV\AA$^6$]&11.3	         &$R_0$ [\AA]&2.67       \\
$\lambda_+$ [\AA$^3$]&0.408      & &\\
$\lambda_-$ [\AA$^3$]&6.431      & &\\
$\rho$ [\AA]&0.3489              & &\\
$\Delta E$ [eV]&2.075            & &\\
\end{tabular}

\caption[x]{\label{tab1}
Parameters for the potentials (\ref{covalent}) and (\ref{ionic}) taken
directly from Ref.\ \cite{engel}.
}
\end{table}

\pagebreak % put figures on separate pages

\pagebreak % put figures on separate pages

\pagebreak % put figures on separate pages

\pagebreak % put figures on separate pages

\end{document}